\documentclass[12pt]{article}
\usepackage{amsmath,amssymb,bm,graphicx}
\usepackage{color} 
\usepackage{hyperref}
\usepackage{slashed}
\usepackage{braket}
\usepackage{caption}
\usepackage{subcaption}
\usepackage{physics}

\usepackage{tikz-feynman}
\tikzfeynmanset{compat=1.1.0}

\usepackage{cite}
\numberwithin{equation}{section}

\newcommand{\secref}[1]{Sec.~\ref{#1}}

\DeclareMathOperator{\diag}{diag}

\begin{document}


\begin{center}
{\Large{\bf Neutrino Flavour Waves\\\vskip 0.3cm Through the Quantum Vacuum: \\\vskip 0.3cm A Theory of Oscillations}}
\end{center}
\vskip .3 truecm
\begin{center}
{\bf { Markku Oksanen, Nico Stirling and Anca Tureanu}}
\end{center}

\begin{center}
\vspace*{0.4cm} 
{\it Department of Physics, University of Helsinki,\\ P.O.Box 64, 
FI-00014 University of Helsinki,
Finland\\
and\\
\vskip 0.1cm
Helsinki Institute of Physics,\\ P.O.Box 64, 
FI-00014 University of Helsinki,
Finland
}
\end{center}
\vspace*{0.2cm} 
\begin{abstract}

We propose a theory for neutrino oscillations, in which the flavour neutrinos are treated as waves of massless particles propagating in a ``refractive quantum vacuum'' and obeying a relativistically covariant equation of motion. The difference in strength between weak interactions and mass-generating interactions is argued to allow for the production and detection of flavour neutrinos in weak interactions as massless particles. They experience the mass-generating interactions as coherent forward scattering in the Brout--Englert--Higgs vacuum, which induces macroscopically multi-refringent effects. The flavour neutrino wave is then found to have a \textit{universal effective refractive mass in vacuum} and a unique group velocity for a given energy. The coherence of the wave is manifest throughout and, at every moment of the propagation, the energy of the waves is the same. The standard oscillation probability in vacuum is obtained and the effects of matter are incorporated in a natural way.
\end{abstract}

\section{Introduction}\label{sec:intro}

The theoretical description of flavour neutrino oscillations was proposed in its standard and most concise form in the work of Gribov and Pontecorvo \cite{Gribov_Pontecorvo}. The formula for the oscillation probability derived in \cite{Gribov_Pontecorvo}, with amendments for three active neutrinos, has been the fundamental theoretical tool for the analysis of experimental data (see, for example, \cite{osc_exp1}). Ever since, numerous ideas and approaches  have been proposed for phenomenological descriptions of neutrino oscillations, compatible with quantum mechanics and quantum field theory. Notwithstanding, the question still remains: how do neutrinos oscillate?

The standard theory of neutrino oscillations and most of the other descriptions are predicated on the assumptions that: 1) the observed flavour neutrino states are coherent superpositions of massive neutrino states of different masses and 2) the waves corresponding to particles of different masses interfere. Both assumptions are essential for justifying the oscillatory patterns of the transition probability from one flavour to another. Nevertheless, both assumptions are in an irreconcilable conflict with quantum mechanics and quantum field theory. This conflict was analyzed in detail in \cite{AT2023} and we shall not go through it here. We recall only that any mixture of states of particles of different masses is by default incoherent \cite{Coleman} in any conceivable version of quantum mechanics or quantum field theory, and this is true whether the particles are described by plane waves or wave trains. In a nutshell, the reason is that the physical states of a system are described by one-dimensional normalized rays in a Hilbert space \cite{BoLoTo} and a superposition of states of different masses does not belong to a Hilbert space. Recently, new arguments have been proposed and debates have arised about the impossibility that antineutrinos emitted in $\beta$-decay be in a coherent superposition of different mass eigenstates (see \cite{SBZ, Cline} and reference therein).

In this paper, we propose a novel theory of neutrino oscillations, in which the real particles, or asymptotic states, are the massless flavour neutrinos of the Standard Model, propagating through a quantum vacuum that acts as a refractive medium that also mixes the flavours. Any mass term in a Lagrangian is the effective expression of an interaction. We will argue that, in the case of neutrinos, due to the feebleness of the mass-generating interaction as compared to the weak interactions, the effect of the mass terms is refractive and not kinematic; in other words, the neutrinos remain massless as participants in weak interactions, but propagate in free space with a definite, calculable speed which is less than $c$.

In our framework, the flavour neutrinos are represented as massless waves, whose quantum mechanical interpretation is that of probability amplitudes. These waves are refracted just as light waves are in a transparent and dispersive medium, by coherent forward scattering. This ensures the coherence needed for explaining the flavour oscillations. Due to the mixing terms in the Lagrangian, the coherent forward scattering can also change one flavour into another, without altering the energy or momentum of the neutrino. This is strongly reminiscent of light birefringence\footnote{Here, we have in mind the {\it circular birefringence}, which takes place when linearly polarized light propagates in an optically active isotropic and homogeneous solution of an enantiomere of a substance with chiral molecules. It is different from the type of birefringence which occurs in anisotropic crystals.}, which was proposed by Wolfenstein \cite{W-biref} as the correct analogy for neutrino oscillations, but it has so far never been explored in an adequate theoretical framework.

An essential aspect of birefringence is, as in any refraction phenomenon, that the frequencies of the incident and emergent waves are always the same, and the refractive index reflects a change in wave length, due to shifts of phase which happen at each scattering site \cite{Foldy, Lax} (see also \cite{Feynman}). Microscopically, the refracted wave is built up as an averaged wave created by coherent forward scattering at each scattering center, which explains why the frequency (equivalent to the energy of neutrinos, in the particle picture) remains the same \cite{Born,Goldberger,Jauch}. For this reason, energy eigenstates, or mass eigenstates, do not appear at all in this formalism. 

Apart from a consistent treatment of flavour neutrinos from production to detection, the present theory has the merit of describing neutrino oscillations as an essentially macroscopic wave process, whose particularities are defined by microscopic interactions. In contrast, the standard analogy by which neutrino oscillations are modeled is with  the quantum mechanical two-level (or multi-level) systems. In the latter case, the oscillations take place among states of localized atoms or molecules, i.e., without spatial propagation. For this reason, the oscillations are customarily treated as a microscopic phenomenon with a periodicity in time, while its space development is brought into play indirectly, with the ad hoc assumption that whatever change happens in time will also propagate in space, because neutrinos move. A paradoxical consequence of the standard treatment is that the speed of propagation of flavour neutrinos, the very ones that oscillate, cannot be defined. In our formulation, however, the speed of propagation of flavour neutrinos of a given energy is perfectly well-defined and unique for all flavours. This is the effect of the existence of a unique refractive mass, which is given by a combination of all the active neutrino mass parameters in the effective Lagrangian.

It has been recently proposed that neutrino oscillations could be caused by coherent interaction of massless flavour neutrinos with dark matter (see, for example, \cite{Choi, AYS_refractive, Ge}). In the present work, we do not introduce such new interactions between neutrinos and matter. However, our scheme enables the incorporation of any new interactions for neutrinos, contributing their effects to the neutrino oscillations and to the universal refractive mass of neutrinos.

We consider the Lagrangian of the active neutrino fields $\nu_\ell(x)$, where $\ell=e,\mu,\tau$ is the flavour index, to which we add an equal number of right-chirality gauge-singlet fields with a corresponding flavour index. The Lagrangian is comprised of the Standard Model (SM) flavour-conserving part and a flavour violating Dirac mass part arising from a Yukawa coupling with the Brout--Englert--Higgs (BEH) scalar field:
\begin{eqnarray}\label{Lagr}
{\cal L}={\cal L}_0+{\cal L}_{mass}+{\cal L}_{CC}+{\cal L}_{NC}.
\end{eqnarray}
Here ${\cal L}_0$ contains the kinetic terms:
\begin{eqnarray}\label{L0}
{\cal L}_0=\bar{\nu}_{\ell L}(x)i\slashed{\partial}\nu_{\ell L}(x) +\bar{\nu}_{\ell R}(x)i\slashed{\partial}\nu_{\ell R}(x),
\end{eqnarray}
${\cal L}_{mass}$ contains mass and mixing terms:
\begin{eqnarray}\label{Lmass}
{\cal L}_{mass}=- \bar \nu_{{\ell}'L}(x)M_{{\ell}'{\ell}}\nu_{{\ell} R}(x)+h.c.,
\end{eqnarray}
where $M_{{\ell}'{\ell}}$ is, in general, a $3\times 3$ complex nondiagonal matrix. Its elements are given by
$$M_{{\ell}'{\ell}}=\frac{v \,y_{{\ell}'{\ell}}}{\sqrt{2}},$$
where $v$ is the vacuum expectation value of the BEH field and $y_{{\ell}'{\ell}}$ are the elements of the complex nondiagonal matrix of Yukawa couplings.
${\cal L}_{CC}$ describes the charged current interactions of neutrinos with the charged lepton fields $\ell(x)$:
\begin{eqnarray}\label{LCC}
{\cal L}_{CC}=-\frac{g}{\sqrt2}\bar{\nu}_{{\ell}L}(x)\gamma_{\mu} \ell_L(x)W^\mu(x)+h.c.,
\end{eqnarray}
and ${\cal L}_{NC}$ describes the neutral current interactions:
\begin{eqnarray}\label{LNC}
{\cal L}_{NC}=-\frac{g}{2\cos{\theta_W}}\bar{\nu}_{{\ell}L}(x)\gamma_{\mu} \nu_{\ell L}(x)Z^\mu(x).
\end{eqnarray}
The mass and mixing terms \eqref{Lmass} are treated as interaction terms, which produce the aforementioned coherent forward scattering in the BEH vacuum.

Although we do not specifically consider the Majorana neutrinos, the scheme works as well for any seesaw Lagrangian which leads to Weinberg's dimension 5 operator:
\begin{eqnarray}
{\cal L}_{Weinberg}=-\frac{1}{ \Lambda}\bar L_{\ell }(x)\tilde H(x)y_{\ell\ell'}\tilde H^T(x)C\bar L^T_{\ell'}(x)+h.c.,
\end{eqnarray}
where $L_{\ell}(x)$ is the SM massless left-handed leptonic doublet of flavour $\ell$, $H(x)$ is the Higgs doublet, $C$ is the charge conjugation matrix, and $\Lambda$ is the seesaw scale (the cutoff scale of the effective theory).

In \secref{sec:flavourwaves} we present our description of neutrino flavour oscillations in the refractive vacuum. The refractive index and oscillation/survival probabilities of flavour neutrinos in the BEH vacuum are found in \secref{sec:refractiveindex}, using the forward scattering amplitude derived in Appendices~\ref{sec:forwardscattering} and \ref{app_phase}. The speed and effective mass of neutrinos are obtained in \secref{sec:speed}. The effect of matter on neutrino oscillations is considered in \secref{sec:matter}. In \secref{sec:masslessness} we argue that for neutrinos the mass-generating interaction is much weaker than the weak interaction, which justifies the assumption that the asymptotic flavour neutrino states are considered to be massless. Finally, the results are discussed in \secref{sec:discussion}.

\section{Flavour oscillations in the quantum vacuum}
\label{sec:flavourwaves}

In this section, we shall develop the theory of flavour oscillations for neutrinos in analogy with the circular birefringence. 
The general framework is the theory of multiple scattering of waves \cite{Lax, Foldy, Born}, which is used to derive the wave equation for the flavour neutrinos. The neutrino waves experience multirefringence by interaction with the background vacuum expectation value of the BEH field; this mixes the flavour components and consequently causes the oscillation of the flavours. We also find that the neutrinos gain a universal effective (refractive) mass and group velocity, which are the same for all flavours. 

Our assumptions are the following:

\begin{enumerate}
\item Every neutrino is produced in weak interactions as a massless particle in a definite flavour state, with definite energy $E$ and momentum $\bf p$, satisfying the dispersion relation $E=|{\bf p}|$. The implicit assumption is that weak interactions strictly conserve flavour, as in the Standard Model. It follows that the free massless neutrino flavour waves are plane waves,
\begin{equation}
\psi_0({\bf x}, t)= e^{-i( E t-{\bf p}\cdot{\bf x})},\ \ \ \  E=|{\bf p}|,
\end{equation}
satisfying the Helmholtz equation
\begin{equation}\label{free_wave}
(\nabla^2+ E^2)\psi_0({\bf x}, t) = 0.
\end{equation}
From the quantum mechanical perspective, this is the Klein--Gordon equation for the massless case,
\begin{equation}\label{KG}
\left(\frac{\partial^2}{\partial t^2}-\Delta\right)\psi_0({\bf x}, t) = 0,
\end{equation}
obtained from the dispersion relation\footnote{We cannot use the linear dispersion relation $E=|{\bf p}|$ to derive the differential equation linear in time derivatives $i\frac{\partial}{\partial t}\psi_0({\bf x}, t)=|-i\nabla|\psi_0({\bf x}, t)$, because the latter gives inconsistent results for the negative energy solutions.} $E^2={\bf p}^2$, with $E=i\frac{\partial}{\partial t}$ and ${\bf p}=-i\nabla$.
Eq. \eqref{free_wave} is compatible with the Weyl equation for massless spinors. $\psi_0({\bf x}, t)$ is the analog of the incident (linearly polarized) wave which enters a medium in the case of circular birefringence.

    \item The mass-generating interactions \eqref{Lmass} are "switched on" and start taking effect after the production of the massless flavour neutrino, inducing  multiple coherent forward scatterings\footnote{In Sect. \ref{sec:masslessness} we argue why this separation between weak interactions and mass generating interactions is plausible.}. We assume an isotropic distribution of scatterers and weakly scattering media, such that the first Born approximation can be applied. The assumption is realistic, due to the homogeneity and spatial uniformity of the background field $v$, and because the interaction of neutrino with the $v$-field is much more feeble then their weak interactions.  The scatterings cause a change of phase as well as a change of flavour, without causing any change in energy -- see Fig. \ref{fig:Fig1}.

\begin{figure}[h]
     \centering
     \begin{subfigure}[b]{0.45\textwidth}
         \centering
         \begin{tikzpicture}
            \begin{feynman}[large]
                \vertex (a);
                \vertex [right=of a] (b);
                \vertex [right=of b] (c);
                \vertex [right=of c] (d);
                \vertex [above=of b] (e);
                \vertex [above=of c] (f);
                \vertex [below=0.2em of b] {\(y^*_{\ell \ell''}\)};
                \vertex [below=0.4em of c] {\(y^{}_{\ell' \ell''}\)};
                
                \diagram* {
                (a) -- [fermion, edge label=\(\nu_{\ell L}\)] (b) -- [fermion, edge label=\(\nu_{\ell'' R}\)] (c)-- [fermion, edge label=\(\nu_{\ell' L}\)] (d),
                (b) -- [scalar, edge label'=\(v\),insertion=1] (e),
                (c) -- [scalar, edge label'=\(v\),insertion=1] (f),
                };
            \end{feynman}
        \end{tikzpicture}
         \caption{}
         \label{fig:y equals x}
     \end{subfigure}
     \begin{subfigure}[b]{0.45\textwidth}
        \centering
        \begin{tikzpicture}
            \begin{feynman}[large]
                \vertex (a);
                \vertex [right=of a] (b);
                \vertex [right=of b] (c);
                \vertex [right=of c] (d);
                \vertex [above=of b] (e);
                \vertex [above=of c] (f);
                \vertex [below=0.4em of b] {\(M^*_{\ell \ell''}\)};
                \vertex [below=0.4em of c] {\(M^{}_{\ell' \ell''}\)};
                
                \diagram* {
                (a) -- [fermion, edge label=\(\nu_{\ell L}\)] (b) -- [fermion,insertion=0.01, insertion=0.99, edge label=\(\nu_{\ell'' R}\)] (c)-- [fermion, edge label=\(\nu_{\ell' L}\)] (d)
                };
            \end{feynman}
        \end{tikzpicture}
        \caption{}
        \label{fig:three sin x}
     \end{subfigure}
    \caption{\small (a) The interaction of the left chiral Weyl neutrino field with the constant scalar field $v$ causes a change of flavour. (b) This is equivalent to mass insertions.}
    \label{fig:Fig1}
\end{figure}
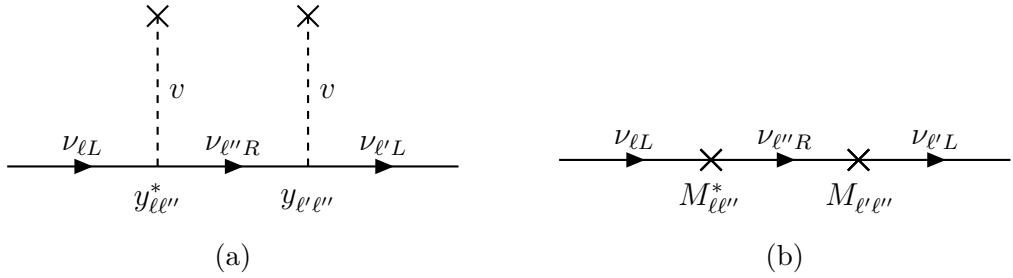

\item Through multiple coherent forward scatterings, the neutrino wave builds into an averaged coherent wave with three flavour components $\Psi_{\nu_\ell}({\bf x},t)$. Each component represents the probability amplitude that the neutrino has a certain flavour at the point $\bf x$ and time $t$. Since only forward coherent scattering is possible in the BEH vacuum, the averaged coherent neutrino wave travels with unchanged energy, or frequency, and without attenuation throughout its macroscopic propagation length.  When the neutrino is detected, it will have the same energy and momentum as when it was created, though the flavour can be different.

\item The averaged coherent neutrino wave  $\Psi_{\nu_{\ell}}({\bf x}, t)$ in the BEH vacuum satisfies the  {\it relativistic wave equation with source}, just like a light wave propagating in a dielectric medium \cite{Born}. In a homogeneous dispersive medium with the number density of scatterers $N$, this is:
\begin{equation}\label{waveequation}
\left[\left(\nabla^2 + E^2\right)\delta_{\ell'\ell}  +4\pi N f_{\ell'\ell}(0)\right]\Psi_{\nu_{\ell}}({\bf x}, t)= 0,
\end{equation}
where $f_{\ell'\ell}(0)$ is the transition amplitude of the forward scattering from flavour $\ell$ to $\ell'$\cite{Lax, Goldberger}.
As explained above, the wave has the definite energy $E$, such that we can separate the variables:
\begin{equation}
\Psi_{\nu_{\ell}}({\bf x}, t)= e^{-i E t}\Psi_{\nu_{\ell}}({\bf x}).
\end{equation}
 If there are several types of scatterers with different scattering amplitudes and densities, formula \eqref{waveequation} correspondingly takes into account all of them separately:
\begin{equation}\label{waveequation_2}
\left[\left(\nabla^2 + E^2\right)\delta_{\ell'\ell}  +4\pi N^\alpha f^\alpha_{\ell'\ell}(0)\right]\Psi_{\nu_{\ell}}({\bf x}, t)= 0,
\end{equation}
where $\alpha$ identifies each type of scatterer \cite{Lax}.
Eq. \eqref{waveequation_2} will be used in Sect. \ref{sec:matter}, where propagation through homogeneous matter will be considered. 

\end{enumerate}

In the rest of this section we shall use these assumptions to find the transition amplitudes and probabilities for the case of purely forward scattering as is the case in the BEH vacuum. For simplicity we consider the case of two flavours, for example $\ell = {e, \mu}$. The results are easily generalised to more flavours at the end of the section. 

With these considerations, we write the averaged coherent wave equation \eqref{waveequation} in the matrix form:
\begin{eqnarray}\label{coupled_eqs}
\left[\nabla^2+ E^2+4\pi N\begin{pmatrix}
            f_{ee}(0) &f_{e\mu}(0)\\
           f_{e\mu}(0)& f_{\mu\mu}(0)
            \end{pmatrix}\right]\begin{pmatrix}
            \Psi_{\nu_e}({\bf x}, t)\\
            \Psi_{\nu_\mu}({\bf x}, t)
            \end{pmatrix}=D_f\begin{pmatrix}
            \Psi_{\nu_e}({\bf x}, t)\\
            \Psi_{\nu_\mu}({\bf x}, t)
            \end{pmatrix}=0,
\end{eqnarray}
with the initial conditions
\begin{eqnarray}
\begin{pmatrix}
            \Psi_{\nu_e}(0,0)\\
            \Psi_{\nu_\mu}(0,0)
            \end{pmatrix}=\begin{pmatrix}
            1\\
            0
            \end{pmatrix}\ \ \ \mbox{or}\ \ \ \begin{pmatrix}
            0\\
            1
            \end{pmatrix},
\end{eqnarray}
since the neutrinos are produced strictly in one of the flavour states. To reiterate, the coherent waves $\Psi_{\nu_{\ell'}}({\bf x}, t)$ represent the probability amplitude to find a neutrino of flavour $\ell'$ at the point $\bf x$ at time $t$, assuming that the neutrino was produced in a given initial state of flavour $\ell$:
\begin{equation}
\Psi_{\nu_{\ell'}}(\mathbf{x},t)= \sum_\ell \mathcal{A}_{\nu_{\ell}\to\nu_{\ell'}}(\mathbf{x},t) \Psi_{\nu_{\ell}}(0,0).
\end{equation}

We diagonalize the differential hermitian operator $D_f$ with the unitary matrix $U$:
\begin{equation}
U^\dagger D_f U= D'_f,\ \ \ \  U=\begin{pmatrix}
            \cos\theta &-\sin\theta\\
           \sin\theta&\cos\theta
            \end{pmatrix},\label{2d-rotation}
\end{equation}
where
\begin{eqnarray}
D'_f=\left[\nabla^2+ E^2+4\pi N\begin{pmatrix}
            f_{1}(0) &0\\
           0& f_{2}(0)
            \end{pmatrix}\right].
\end{eqnarray}
Since we discuss two-flavour Dirac neutrino mixing, it is sufficient to take $U$ as a rotation matrix. The general case will be covered later, when we consider an arbitrary number of flavours.

In the new basis,
\begin{equation}\label{change_basis}
U^\dagger \begin{pmatrix}
            \Psi_{\nu_e}({\bf x}, t)\\
            \Psi_{\nu_\mu}({\bf x}, t)
            \end{pmatrix}
            =\begin{pmatrix}
            \Psi_{1}({\bf x}, t)\\
            \Psi_{2}({\bf x}, t)
            \end{pmatrix},
\end{equation}
we get the decoupled wave equations
\begin{eqnarray}\label{decoupled}
\left[\nabla^2+ E^2+4\pi N\begin{pmatrix}
            f_{1}(0) &0\\
           0& f_{2}(0)
            \end{pmatrix}\right]\begin{pmatrix}
            \Psi_{1}({\bf x}, t)\\
            \Psi_{2}({\bf x}, t)
            \end{pmatrix}=0,
\end{eqnarray}
with the obvious plane-wave solution
\begin{equation}\label{eigenfunction}
\Psi_i({\bf x}, t)=  e^{-i( E t-{\bf p}'_i\cdot{\bf x})},\ \ \ i=1,2,
\end{equation}
where
\begin{equation}
{\bf p}'^2_i=  
{\bf p}^2+4\pi N f_i(0) ,\ \ \  E^2={\bf p}^2,\ \ \ i=1,2.
\end{equation}
We emphasize that $f_i(0)$ do not represent forward scattering amplitudes of any physical processes and, as a matter of fact, do not have a physical meaning. They are just notations for the eigenvalues of the nondiagonal matrix of forward scattering amplitudes\footnote{In the case of the interaction with the BEH vacuum, discussed in this work, the eigenvalues $f_i(0)$ are negative. In general, when other possible interactions with matter are considered, $f_i(0)$ can be positive or even complex, when there is absorption.} that appear in \eqref{coupled_eqs}. Consequently, we will not attach any physical meaning either to the waves $\Psi_i({\bf x}, t)$.

We define the refractive indices
\begin{equation}
n_i^2=\frac{{\bf p}'^2_i}{{\bf p}^2}
=1+4\pi \frac{N f_i(0)}{ E^2} ,\ \ \ i=1,2.
\end{equation}
In the limit $\frac{N f_i(0)}{ E^2}\ll1$,
\begin{equation}
n_i=1+2\pi \frac{N f_i(0)}{ E^2} ,\ \ \ i=1,2.
\end{equation}
The flavour neutrino wave is, using \eqref{change_basis} and \eqref{eigenfunction},
\begin{align}
\begin{pmatrix}
            \Psi_{\nu_e}({\bf x})\\
            \Psi_{\nu_\mu}({\bf x})
            \end{pmatrix}&=U\begin{pmatrix}
            \Psi_{1}({\bf x})\\
            \Psi_{2}({\bf x})
            \end{pmatrix}\notag\\
            &=U\begin{pmatrix}
           e^{in_1{\bf p}\cdot({\bf x}-{\bf x}_0)} &0\\
           0& e^{in_2{\bf p}\cdot({\bf x}-{\bf x}_0)}
            \end{pmatrix}
            U^{\dagger}\begin{pmatrix}
            \Psi_{\nu_e}({\bf x}_0)\\
            \Psi_{\nu_\mu}({\bf x}_0)
            \end{pmatrix},
\end{align}
namely
\begin{equation}
\begin{pmatrix}
            \Psi_{\nu_e}({\bf x})\\
            \Psi_{\nu_\mu}({\bf x})
            \end{pmatrix}=\begin{pmatrix}
           e^{in_1{\bf p}\cdot{\bf x}}\cos^2\theta+ e^{in_2{\bf p}\cdot{\bf x}}\sin^2\theta&(e^{in_1{\bf p}\cdot{\bf x}}-e^{in_2{\bf p}\cdot{\bf x}})\sin\theta\cos\theta\\
           (e^{in_1{\bf p}\cdot{\bf x}}-e^{in_2{\bf p}\cdot{\bf x}})\sin\theta\cos\theta& e^{in_1{\bf p}\cdot{\bf x}}\sin^2\theta+ e^{in_2{\bf p}\cdot{\bf x}}\cos^2\theta
            \end{pmatrix}
\begin{pmatrix}
            \Psi_{\nu_e}(0)\\
            \Psi_{\nu_\mu}(0)
            \end{pmatrix}.
\end{equation}

Let us consider that an electron neutrino was produced at $ {\bf x}=0, t=0$. Then, our initial condition is
\begin{equation}\label{initialcondition.electron}
\begin{pmatrix}
            \Psi_{\nu_e}(0,0)\\
            \Psi_{\nu_\mu}(0,0)
            \end{pmatrix}=\begin{pmatrix}
            1\\
            0
            \end{pmatrix}.
\end{equation}
With this initial condition, we find the transition amplitudes to be:
\begin{align}
           \begin{pmatrix}
            {\cal A}_{\nu_e\to\nu_e}({\bf x},t)\\
             {\cal A}_{\nu_e\to \nu_\mu}({\bf x},t)
            \end{pmatrix}
            &=\begin{pmatrix} \Psi_{\nu_e}({\bf x},t)\\
            \Psi_{\nu_\mu}({\bf x},t)
            \end{pmatrix}\notag\\
            &=e^{-i E t}\begin{pmatrix}
            e^{in_1{\bf p}\cdot{\bf x}}\cos^2\theta+ e^{in_2{\bf p}\cdot{\bf x}}\sin^2\theta\\
            (e^{in_1{\bf p}\cdot{\bf x}}-e^{in_2{\bf p}\cdot{\bf x}})\sin\theta\cos\theta
            \end{pmatrix}\notag\\
            &=e^{-i( E t-\bar n\,{\bf p}\cdot{\bf x})}\begin{pmatrix}
            \cos \left(\frac{\Delta n}{2}{\bf p}\cdot{\bf x} \right)+i\cos 2\theta\, \sin\left(\frac{\Delta n}{2}{\bf p}\cdot{\bf x}\right)\\
            i\sin2\theta\,\sin \left(\frac{\Delta n}{2}{\bf p}\cdot{\bf x}\right)
            \end{pmatrix},
            \label{transitionamplitudes}
\end{align}
where $\bar n=\frac{n_1+n_2}{2}$ and $\Delta n=n_1-n_2$.

The survival and transition probabilities at a distance $L=|\mathbf{x}|$ from production, in the direction of propagation, $\mathbf{x}=L\frac{{\bf p}}{|{\bf p}|}$, are
\begin{align}
    P_{\nu_e\to\nu_e}(L, E)&=\left|{\cal A}_{\nu_e\to\nu_e}(L,t)\right|^2
    =1-\sin^22\theta\sin^2\left(\frac{\Delta n}{2} E L\right),\\
    P_{\nu_e\to\nu_\mu}(L, E)&=\left|{\cal A}_{\nu_e\to\nu_\mu}(L,t)\right|^2
    =\sin^22\theta\sin^2\left(\frac{\Delta n}{2} E L\right),
\end{align}
where $ E=|\mathbf{p}|$ is the energy of the neutrino.

 It is now easy to see how this scheme is formally similar to the circular birefringence: the flavour neutrino waves $\Psi_{\nu_\ell}({\bf x},t),\ \ell=e,\mu$, are the analogues of two orthogonal linear polarization, of which one enters the medium. The optically active medium mixes the two linear polarizations through the coherent scatterings that take place inside it. The waves $\Psi_{i}({\bf x},t),\ i=1,2$ are the analogues of the left- and right-circular polarizations, which have definite but different refraction indices, $n_L$ and $n_R$. As a result, their optical paths corresponding to a certain traveled length $L$ are different and, when light exists the medium, it is still linearly polarized, but on a different direction. The angle of rotation of the polarization plane is given by $\Delta \phi=\Delta n L E/2$, with $\Delta n=n_L-n_R$. 

In the case of circular birefringence, the stereoisomers composing the medium are chiral molecules and the phenomenon reflects the parity breaking; in the same way, neutrino oscillations reflect the flavour violating interactions. Although the vacuum does not violate flavour, it is nevertheless blind to flavour, because the flavour quantum number does not determine a superselection rule in the Hilbert space of massless neutrinos.

\paragraph{Generalization to more flavours:} We consider now $F>2$ flavours mixing. The two-dimensional rotation matrix in \eqref{2d-rotation} is replaced by a unitary $F\times F$ matrix $U$ with components $U_{\ell i}$.
The averaged coherent neutrino waves are obtained as
\begin{equation}\label{neutrinowave}
\begin{split}
\Psi_{\nu_{\ell'}}(\mathbf{x},t)&=\sum_{i}\sum_{\ell}U_{\ell' i}U^{*}_{\ell i}
\,e^{-i( E t-n_i\mathbf{p}\cdot\mathbf{x})}\,\Psi_{\nu_\ell}(0,0)\\
&=e^{-i( E t-\bar{n}\mathbf{p}\cdot\mathbf{x})}
\sum_{i}\sum_{\ell}U_{\ell' i}U^{*}_{\ell i}
\,e^{\frac{i}{F}\sum_{j\neq i}\Delta n_{ij}\mathbf{p}\cdot\mathbf{x}}
\,\Psi_{\nu_\ell}(0,0)
\end{split}
\end{equation}
where $\bar n=\frac{1}{F}\sum_{i=1}^Fn_i$ and $\Delta n_{ij}=n_i-n_j$.
The amplitude for a neutrino $\nu_{\ell}$ at $(\mathbf{x}=0,t=0)$ to transition to a neutrino $\nu_{\ell'}$ at $(\mathbf{x},t)$ is
\begin{equation}\label{gen_amplitude}
\mathcal{A}_{\nu_{\ell}\to\nu_{\ell'}}(\mathbf{x},t)
=e^{-i( E t-\bar{n}\mathbf{p}\cdot\mathbf{x})}\sum_{i}U_{\ell' i}U^{*}_{\ell i}
\,e^{\frac{i}{F}\sum_{j\neq i}\Delta n_{ij}\mathbf{p}\cdot\mathbf{x}}.
\end{equation}
Noting that 
\begin{equation}
\frac{1}{F}\left(\sum_{k\neq i}\Delta n_{ik}-\sum_{l\neq j}\Delta n_{jl}\right) =\Delta n_{ij},
\end{equation}
the transition probability at the distance $L$ in the direction of propagation is found to be
\begin{equation}\label{transitionprobability}
P_{\nu_{\ell}\to\nu_{\ell'}}(L, E)=
\left|\mathcal{A}_{\nu_{\ell}\to\nu_{\ell'}}(L,t)\right|^2
=\sum_{i,j}U_{\ell' i}U^{*}_{\ell' j}U^{*}_{\ell i}U_{\ell j}\,e^{i\Delta n_{ij} E L}.
\end{equation}

\section{Refractive indices and flavour oscillation probabilities in the BEH vacuum}\label{sec:refractiveindex}

In the theory of multiple scattering of waves, the scattering amplitude $f_{\ell' \ell}(0)$ for the forward scattering process $\nu_{\ell L}\to\nu_{\ell'L}$ is related to the differential cross section of interaction by the formula
\begin{equation}\label{scatteringamplitude'}
    |f_{\ell' \ell}(0)| = \sqrt{\frac{d \sigma_{\ell \rightarrow\ell'}}{d \Omega}\Big|_{\theta,\phi=0}},
\end{equation}
which is the well-known relation from quantum mechanical scattering theory.
The forward scattering amplitudes arising from the mass-generating interactions in the BEH vacuum are calculated in Appendix \ref{sec:forwardscattering}. Using eq. \eqref{forward_ampl}, we obtain
\begin{equation}\label{f(0)-matrix}
    4\pi N f_{\ell' \ell}(0)=-2\left( M M^\dagger \right)_{\ell'\ell},
\end{equation}
where we have taken the number density of scatterers as $N=\frac{1}{V}$, in agreement with the definition of the normalization volume. Then, the flavour neutrino waves will satisfy the relativistic wave equation \eqref{waveequation_2} with invariant scattering term\footnote{We emphasize that the scattering cross-section given by eq. \eqref{differentialcrosssection} and the density of scatterers $N=1/V$ are not observables quantities, because they both depend on the normalization volume. The observable of this formalism is the product given by \eqref{f(0)-matrix}. This quantity ultimately leads to the phase shift and the measurable refractive index.},
\begin{equation}\label{waveequation_BEH}
\left[\left(\nabla^2 + E^2\right)\delta_{\ell'\ell} -2\left( M M^\dagger \right)_{\ell'\ell}\right]\Psi_{\nu_{\ell}}({\bf x}, t)= 0.
\end{equation}

The complex nonsingular mass matrix $M$ (with the notation in \eqref{Lmass}) is diagonalized by a biunitary transformation:
\begin{equation}\label{M_diag}
U^{\dagger}MV=M_\mathrm{diag}=\frac{1}{\sqrt{2}}\diag(m_1,m_2,\ldots,m_F),
\end{equation}
therefore $$U^{\dagger}MM^{\dagger}U=\frac{1}{2}\diag(m_1^2,m_2^2,\ldots,m_F^2).$$ 
The scaling factor $\frac{1}{\sqrt{2}}$ in \eqref{M_diag} is included so that $m^2_i$, $i=1,2,\ldots,F$, are the eigenvalues of the matrix $2MM^\dagger$ in \eqref{f(0)-matrix}\footnote{The parameters $m_i$ do not represent masses of particles in our approach. We adopt this notation because it is consacrated in the literature and it makes the comparison with known results straightforward.}.
The decoupled waves $\Psi_i(\mathbf{x},t)=U^{*}_{\ell i}\Psi_{\nu_\ell}(\mathbf{x},t)$ satisfy the wave equations \eqref{decoupled}, with
\begin{equation}
    4\pi N f_i(0)=-m_i^2,\quad i=1,2,\ldots,F.
\end{equation}
Hence, the eigenvalues of the refractive indices are
\begin{equation}\label{refractiveindices}
n_i^2=1-\frac{m_i^2}{ E^2},\quad i=1,2,\ldots,F.
\end{equation}
The refractive index that defines the propagation velocity of the flavour neutrino wave \eqref{neutrinowave} is the arithmetic mean $\bar n$ of $n_i$.
For neutrinos with energies satisfying $ E^2\gg \overline {m^2}$,
\begin{equation}\label{refractiveindices.highenergy}
    \bar n=1-\frac{\overline{m^2}}{2 E^2},\qquad
    \overline{m^2}=\frac{1}{F}\sum_{i=1}^Fm_i^2.
\end{equation}
 
\begin{figure}[h]
    \centering
    \includegraphics[width=16cm]{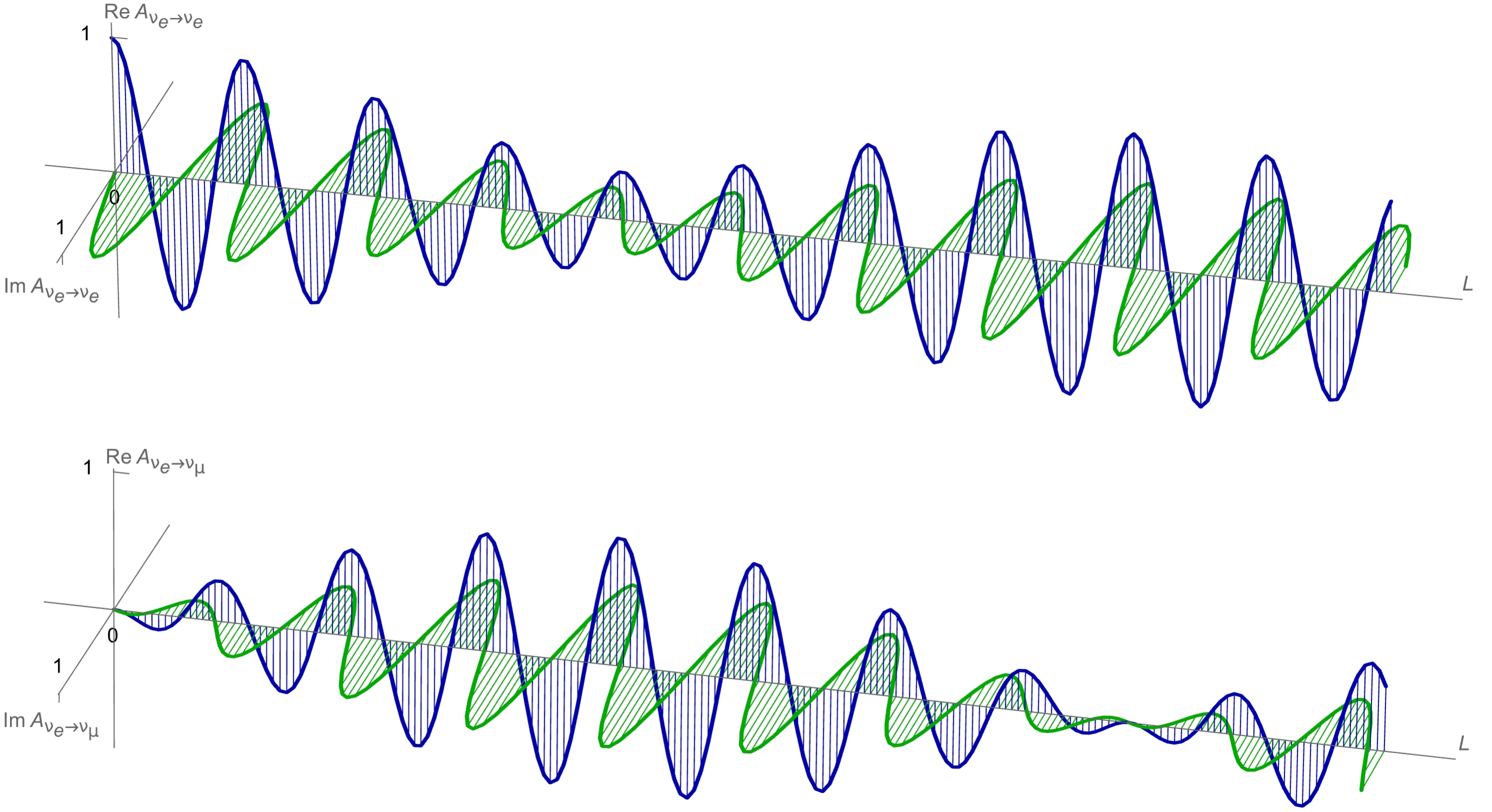}
    
        \caption{\small An illustration of two flavour oscillations, given by eqs. \eqref{transitionamplitudes}, with $\bar n=1-\frac{m_1^2+m_2^2}{4E^2}$ and $\Delta n=(m_2^2-m_1^2)/2E^2$. The real and imaginary components (blue and green, respectively) of the survival and transition amplitudes (top and bottom, respectively) for an electron neutrino initially at $L = 0$ and $t = 0$, as a function of the distance $L$, are shown. The following numerical values are used: $ E=10\,\rm{eV}$, $m_1 = 0 \,\rm{eV}$, $m_2=5\,\rm{eV}$ and $\theta = 0.6$ radians. An energy that is much smaller than in usual experiments is used for visual reasons. For realistic energies the wavelength of the underling wave is much smaller than the length of modulation.}
    \label{fig:flavour-wave}
\end{figure}

The amplitudes \eqref{gen_amplitude} of the flavour waves \eqref{neutrinowave}, 
\begin{equation*}
\Psi_{\nu_{\ell'}}(\mathbf{x},t)= \sum_\ell \mathcal{A}_{\nu_{\ell}\to\nu_{\ell'}}(\mathbf{x},t) \Psi_{\nu_{\ell}}(0,0),
\end{equation*}
for neutrinos with the refractive index \eqref{refractiveindices.highenergy} are
\begin{equation}
\mathcal{A}_{\nu_{\ell}\to\nu_{\ell'}}(\mathbf{x},t)
=e^{-i( E t-\bar{n}\,\mathbf{p}\cdot\mathbf{x})}\sum_{i}U_{\ell' i}U^{*}_{\ell i}\,
e^{-\frac{i}{F}\sum_{j\neq i}\frac{\Delta m^2_{ij}}{2 E^2}\mathbf{p}\cdot\mathbf{x}},
\end{equation}
where $\Delta m_{ij}^2=m_i^2-m_j^2$. For two flavours mixing, with the initial condition \eqref{initialcondition.electron}, the transition and survival amplitudes are illustrated in Fig.~\ref{fig:flavour-wave}.

The flavour transition probabilities \eqref{transitionprobability} become:
\begin{equation}\label{oscillationprobability}
    P_{\nu_{\ell}\to\nu_{\ell'}}(L, E)
    =\sum_{i,j}U_{\ell' i}U^{*}_{\ell' j}U^{*}_{\ell i}U_{\ell j} \exp\left(-i\frac{\Delta m_{ij}^2L}{2 E}\right),
\end{equation}
or, in a form which is more convenient for studying the CP properties,
\begin{eqnarray}\label{oscillationprobability2}
    P_{\nu_{\ell}\to\nu_{\ell'}}(L, E)
    &=&\delta_{\ell\ell'}-
    4\Re \sum_{i>j}U_{\ell' i}U^{*}_{\ell' j}U^{*}_{\ell i}U_{\ell j} \sin^2\left(\frac{\Delta m_{ij}^2L}{4 E}\right)\cr
   &+&2\Im \sum_{i>j}U_{\ell' i}U^{*}_{\ell' j}U^{*}_{\ell i}U_{\ell j} \sin\left(\frac{\Delta m_{ij}^2L}{ E}\right).
\end{eqnarray}

Antineutrinos have the same refractive indices, as can be easily seen from formula \eqref{forward_ampl}. The decoupled antineutrino wave modes are defined by 
\begin{equation}
\Psi_{i}(\mathbf{x},t)=U_{\ell i}\Psi_{\bar\nu_\ell}(\mathbf{x},t),
\end{equation}
since the forward amplitude \eqref{forward_ampl} for antineutrinos is diagonalized by $U^{*}$.
For antineutrinos the transition probability is obtained by replacing $U$ with $U^{*}$ in \eqref{oscillationprobability},
\begin{equation}
    P_{\bar\nu_{\ell}\to\bar\nu_{\ell'}}(L, E)
    =\sum_{i,j}U^{*}_{\ell' i}U_{\ell' j}U_{\ell i}U^{*}_{\ell j} \exp\left(-i\frac{\Delta m_{ij}^2L}{2 E}\right).
\end{equation}
When putting the above formula in a form similar to \eqref{oscillationprobability2}, the last term containing imaginary part gets a negative sign.
The probabilities for the transitions $\bar\nu_{\ell}\to\bar\nu_{\ell'}$ and $\nu_{\ell}\to\nu_{\ell'}$ are equal, $P_{\bar\nu_{\ell}\to\bar\nu_{\ell'}}=P_{\nu_{\ell}\to\nu_{\ell'}}$, when the imaginary parts are zero, but they are different from each other $P_{\bar\nu_{\ell}\to\bar\nu_{\ell'}}\neq P_{\nu_{\ell}\to\nu_{\ell'}}$ when the mixing matrix contains phases that cannot be factorized, which is the case when CP is violated. The whole analysis of CP violation proceeds from here in the standard way.

The probabilities for the transitions $\bar\nu_{\ell}\to\bar\nu_{\ell'}$ and $\nu_{\ell'}\to\nu_{\ell}$ are equal, $P_{\bar\nu_{\ell}\to\bar\nu_{\ell'}}=P_{\nu_{\ell'}\to\nu_{\ell}}$, which is a consequence of the CPT invariance.

\section{Speed of the neutrino waves and effective mass of flavour neutrinos}
\label{sec:speed}

\begin{figure}[h]
    \centering
    \includegraphics[width=14cm]{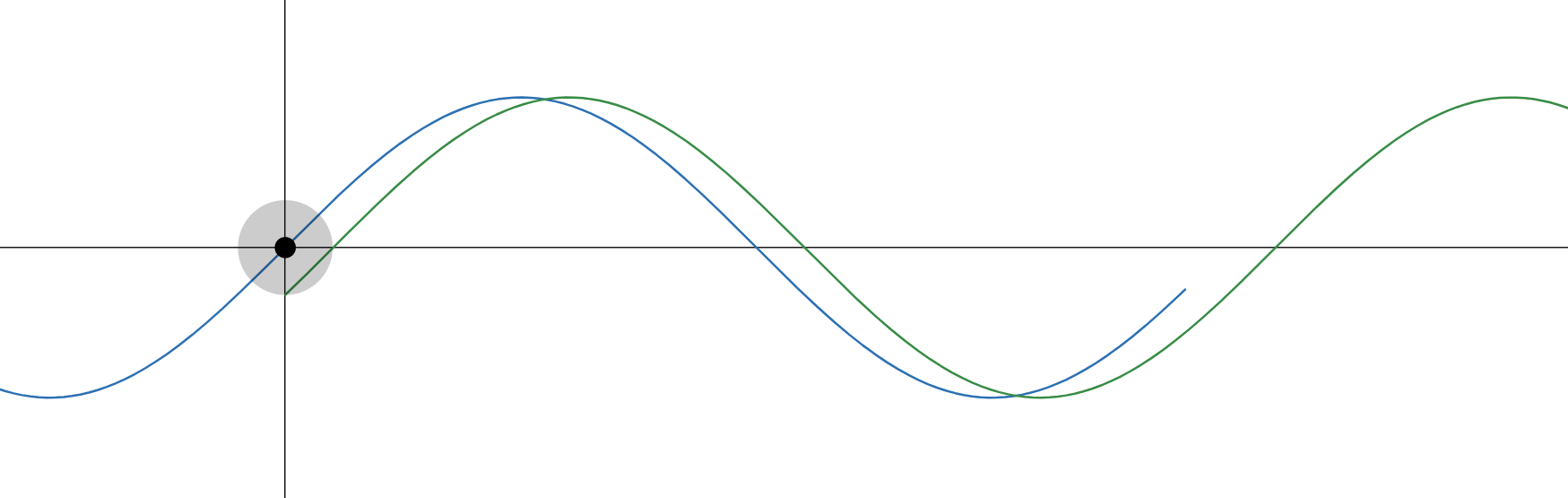}
    
        \caption{\small Illustration of a leading scattered wave. The incoming blue wave scatters from the black dot. The scattering causes a negative shift in phase, but without a change in frequency, namely the green wave. The scattered wave is leading and the phase seems to travel faster than the speed of light. The actual speed of the wave formed by the superposition of the incident and emergent waves is the group velocity.}
    \label{fig:leadingwave}
\end{figure}

Eq. \eqref{gen_amplitude} shows that all the transition/survival amplitudes $\mathcal{A}_{\nu_{\ell}\to\nu_{\ell'}}(\mathbf{x},t)$ have the same space-time evolution, governed by the phase
\begin{equation}
e^{-i( E t-\mathbf{p}'\cdot\mathbf{x})}=e^{-i( E t-\bar{n}\,\mathbf{p}\cdot\mathbf{x})},    
\end{equation}
to which all the refraction index eigenvalues contribute through the average $\bar n$. If we consider the medium to be the BEH vacuum, then $\bar n$ is given by \eqref{refractiveindices.highenergy} when $E^2\gg \overline{m^2}$, and the energy dependence indicates that the medium is dispersive. As the refraction index is subunitary, the phase velocity is greater than the speed of light in vacuum, $v_p>c$. This is the result of the negative phase difference induced by the interaction with the BEH vacuum, which makes the forward-scattered wave advanced (leading) with respect to the incident wave, as depicted in Fig. \ref{fig:leadingwave}. The coherence throughout propagation is automatically ensured.

The speed of the neutrino waves is the group velocity, $v_g=\partial  E/\partial p'$. Noting that $p'=\bar n( E)\,p=\bar n( E)\, E$, we find:
\begin{equation}\label{vg}
v_g=\left(\frac{\partial p'}{\partial  E}\right)^{-1}
=\frac{1}{\bar n( E)+ E\frac{\partial \bar n( E)}{\partial E}} 
=  \frac{1}{1+\frac{\overline {m^2}}{2 E^2}}<1. 
\end{equation}

We may define a so-called ``refractive mass'' of the flavour neutrinos, by comparing the dispersion relation of neutrinos:
$$p'= E\left(1-\frac{\overline{m^2}}{2 E^2}\right)$$
with the dispersion relation of a ultrarelativistic particle of mass $m$ and energy $E$:
$$p=E\left(1-\frac{m^2}{2E^2}\right).$$
The refractive mass of flavour neutrinos is then
\begin{equation}\label{refractivemass}
m_{refr}^2=\overline{m^2}.    
\end{equation}
This is a {\it universal effective mass} for all flavour neutrinos in vacuum, corresponding to a universal speed for all the neutrinos of a given energy \eqref{vg}.

\paragraph{Low-energy neutrinos:} The framework that we have proposed works reliably when the interaction can be treated as perturbation. In the following, we shall explore the obtained formulas in the low-energy regime, with the understanding that at least the general features will survive in a more careful treatment. 

For neutrinos with energies comparable to or below $\overline{m^2}$, we have to use the average refraction index according to the formula \eqref{refractiveindices}, namely,
\begin{equation}\label{refractiveindices.lowenergy}
    \bar n=\frac{1}{F}\sum_{i=1}^F\left(1-\frac{{m_i^2}}{ E^2}\right)^{1/2}.
\end{equation}
The group velocity is
\begin{equation}
v_g=\frac{1}{\frac{1}{F}\sum\limits_{i=1}^F\left(1-\frac{{m_i^2}}{E^2}\right)^{-1/2}},
\end{equation}
where the denominator is the arithmetic mean of $\frac{1}{n_i}$.

When the neutrinos are emitted with energy $E< {\rm min}\,(m_1,m_2,\ldots, m_F)$, the refractive index is purely imaginary, the neutrino wave becomes evanescent and the neutrinos do not propagate. Their "penetration depth", i.e. the distance after which the probability to find the neutrino drops by $1/e$ is $\delta=\frac{1}{2E|\overline {n}(E)|}$. 

If the energy $E$ is in between some mass eigenstates, for example $m_1<E<m_2$, the refractive index is complex, and the probability that the wave will propagate becomes subunitary.

The evanescence of low-energy neutrinos is analogous to the evanescence of electromagnetic waves in plasma, when their frequency is less than the plasma frequency. In our quantum mechanical scheme, this feature is connected with tunneling through a potential barrier, which is in this case produced by the BEH vacuum. 

\section{Oscillations in matter of constant density}\label{sec:matter}

Here we shall sketch the case of oscillations in matter, which follows broadly the vacuum case described above. For definiteness, we shall consider two-flavour mixing and the effect of matter made up of electrons, protons and neutrons. The number densities of the matter constituents is assumed to be constant and the number densities of electrons and protons are assumed to be equal. The neutrinos interact with the matter through the charged-current (CC) and neutral-current (NC) weak interactions and it is assumed that the neutrino energy is much below the mass of the $W$ boson.

Now with multiple types of scatterers we have to use the wave equation with multiple sources \eqref{waveequation_2}. The wave equation for $\Psi_{\nu_\ell}$, where $ \ell=e, \mu$, will read:
\begin{equation}\label{coupled_eqs_m}
\left[\nabla^2+ E^2+4\pi \begin{pmatrix}
            Nf_{ee}(0)+N_e f_{ee}^{e,CC}(0)+N_n f_{ee}^{n,NC}(0)&Nf_{e\mu}(0)\\
           Nf_{e\mu}(0)&N f_{\mu\mu}(0)+N_n f_{\mu\mu}^{n,NC}(0)
            \end{pmatrix}\right]\begin{pmatrix}
            \Psi_{\nu_e}({\bf x}, t)\\
            \Psi_{\nu_\mu}({\bf x}, t)
            \end{pmatrix}=0,
\end{equation}
where $f_{\ell\ell'}(0)$ are given by \eqref{forward_ampl}, $N=1/V$, $N_e$ is the density of electrons, $N_n$ is the density of neutrons, and 
\begin{equation}\label{forward_ampl.matter}
f_{ee}^{e,CC}(0)=-\frac{1}{\sqrt 2\pi}G_F E,\qquad
f_{ee}^{n,NC}(0)=f_{\mu\mu}^{n,NC}(0)=\frac{1}{2\sqrt 2\pi}G_F E,
\end{equation}
are the charged-current and neutral-current forward scattering amplitudes of the processes $\nu_e+e\to\nu_e+e$ and $\nu_\ell+n\to\nu_\ell+n$, respectively, in the low-energy limit (Fermi's interaction approximation), $E\ll M_W$ \cite{Langacker}.
The NC contributions from electrons and protons cancel each other, since the densities of electrons and protons are equal and the NC forward scattering amplitudes are opposite.

Assuming that electron neutrinos of energy $E$ are produced at ${\bf x}=0, t=0$, and the density of electrons is constant, the transition amplitudes become:
\begin{align}
           \begin{pmatrix}
            {\cal A}_{\nu_e\to\nu_e}({\bf x},t)\\
             {\cal A}_{\nu_e\to \nu_\mu}({\bf x},t)
            \end{pmatrix}&=\begin{pmatrix} \Psi_{\nu_e}({\bf x},t)\\
            \Psi_{\nu_\mu}({\bf x},t)
            \end{pmatrix}\notag\\
            &=e^{-i( E t-\bar n\,{\bf p}\cdot{\bf x})}\begin{pmatrix}
            \cos\left( \frac{\Delta n}{2}{\bf p}\cdot{\bf x}\right) +i\cos 2\theta_m \sin\left(\frac{\Delta n}{2}{\bf p}\cdot{\bf x}\right)\\
            i\sin2\theta_m\sin\left( \frac{\Delta n}{2}{\bf p}\cdot{\bf x}\right)
            \end{pmatrix},
\end{align}
where 
\begin{align}\label{medium_parameters}
\bar n&=\frac{n_1+n_2}{2},\qquad \Delta n=n_1-n_2,\notag\\
n_{1,2}^2&=1-\frac{1}{2E^2}\left[m_1^2+m_2^2+A_{CC}-A_{NC}\mp\sqrt{(\Delta m^2\cos2\theta-A_{CC})^2+(\Delta m^2\sin\theta)^2}\right],\notag\\
A_{CC}&= 2\sqrt 2 G_F N_eE,\qquad A_{NC}= 2\sqrt 2 G_F N_nE,\qquad \Delta m^2=m_2^2-m_1^2,\notag\\
\tan 2\theta_m&=\frac{\Delta m^2\sin 2\theta}{\Delta m^2 \cos 2\theta -A_{CC}}.
\end{align}
For sufficiently high energy neutrinos, such that $E\gg m_i$, $E\gg G_F N_e$ and $E\gg G_F N_n$,
\begin{equation}
\bar n= 1-\frac{m_1^2+m_2^2}{4E^2}-\frac{G_F (N_e-N_n)}{\sqrt{2}E},
\end{equation}
the refractive mass of neutrinos becomes
\begin{equation}\label{refractivemass.matter}
m_{refr}^2=\overline{m^2}+\sqrt 2 G_F (N_e-N_n) E,
\end{equation}
while the group velocity does not receive any additional contribution due to matter,
\begin{equation}
v_g=\frac{1}{\bar n( E)+ E\frac{\partial \bar n( E)}{\partial E}} 
=  \frac{1}{1+\frac{\overline {m^2}}{2 E^2}}<1. 
\end{equation}
Once more, the energy of the flavour neutrinos during propagation in matter is the energy with which they are produced at the vertex; the introduction of "energy eigenstates in matter" is not necessary. Nevertheless, the mixing angle in matter is modified and the phenomenon of resonant enhancement of oscillations is also present for neutrinos, but not for antineutrinos, which have opposite sign for the transition amplitude in matter compared to \eqref{forward_ampl.matter}. The evanescence for low-energy neutrinos occurs as well, but the cutoff energies depend on the properties of the medium, as can be seen from \eqref{medium_parameters}.

\section{Masslessness of asymptotic flavour neutrino states}
\label{sec:masslessness}

A key assumption of the present theory of flavour neutrinos is that the asymptotic flavour neutrino states are massless. These are the only physical states of the theory. The effective mass of the propagating flavour neutrinos appears due to their interaction with the vacuum BEH field, which leads to the refractive index of the vacuum (as given in \secref{sec:refractiveindex}). Here we aim to establish that the flavour neutrinos are created and annihilated in weak interactions as massless particles by arguing that the mass-generating interactions described by the Lagrangian \eqref{Lmass} are much more feeble than the weak interactions that create and annihilate the neutrinos. 

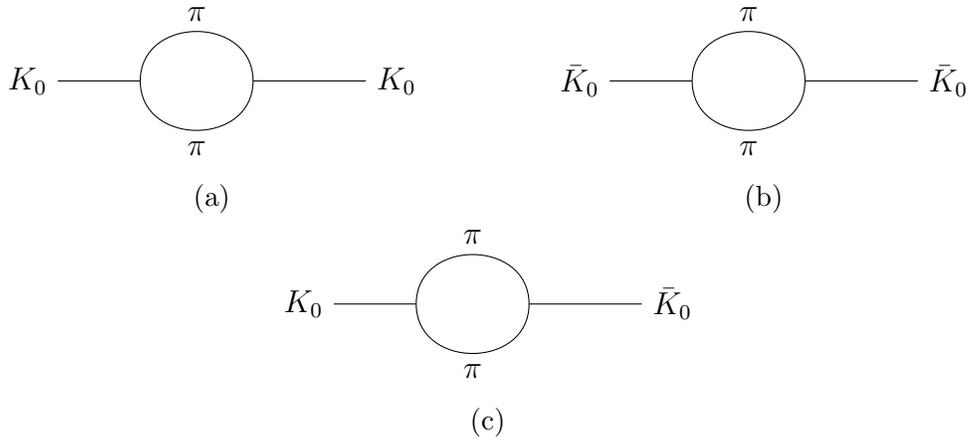
\begin{figure}[h]
    \centering
    \begin{subfigure}[b]{0.45\textwidth}
    \centering
        \begin{tikzpicture}
            \begin{feynman}
                \vertex (a){\(K_0\)};
                \vertex [right=of a] (b);
                \vertex [right=of b] (c);
                \vertex [right=of c] (d){\(K_0\)};
                
                \diagram* {
                (a) --  (b) -- [half left, looseness=1.5, edge label=\(\pi\)] (c)-- [half left, looseness=1.5, edge label=\(\pi\)] (b), (c) -- (d)
            };
            \end{feynman}
        \end{tikzpicture}
        \caption{}
    \end{subfigure}
    \begin{subfigure}[b]{0.45\textwidth}
    \centering
        \begin{tikzpicture}
            \begin{feynman}
                \vertex (a){\(\bar K_0\)};
                \vertex [right=of a] (b);
                \vertex [right=of b] (c);
                \vertex [right=of c] (d){\(\bar K_0\)};

                \diagram* {
                (a) --  (b) -- [half left, looseness=1.5, edge label=\(\pi\)] (c)-- [half left, looseness=1.5, edge label=\(\pi\)] (b), (c) -- (d)
            };
            \end{feynman}
        \end{tikzpicture}
        \caption{}
    \end{subfigure}
        
    \begin{subfigure}[b]{0.45\textwidth}
    \centering
        \begin{tikzpicture}
            \begin{feynman}
                \vertex (a){\(K_0\)};
                \vertex [right=of a] (b);
                \vertex [right=of b] (c);
                \vertex [right=of c] (d){\(\bar K_0\)};

                \diagram* {
                (a) --  (b) -- [half left, looseness=1.5, edge label=\(\pi\)] (c)-- [half left, looseness=1.5, edge label=\(\pi\)] (b), (c) -- (d)
            };
            \end{feynman}
        \end{tikzpicture}
        \caption{}
        \label{fig:kaons}
    \end{subfigure}
    \caption{\small The virtual weak interaction processes by which a $K_0$ beam creates $\bar K_0$ in a coherent manner, {\it after} $K_0$ had been produced in strong interactions.}
\end{figure}

We argue this point in analogy with the standard theory of $K_0-\bar K_0$ mixing and oscillations (for a review, see \cite{Nierste}): the $K_0$ mesons are produced by strong interaction, neglecting the strangeness-violating effect of the weak interaction \cite{GMP, PP}. Insofar as weak interactions are neglected, $K_0$ and $\bar K_0$ are completely stable mass-degenerate particles\footnote{For this reason, the measurements of the mass difference between $K_0$ and $\bar K_0$ are used for imposing bounds on CPT violation.} \cite{Nishijima_fundamental}. The weak interaction takes effect after a $K_0$ has been created, producing coherently, by virtual processes, the antiparticle $\bar K_0$ (see Fig. \ref{fig:kaons}). The same weak interactions lead to the appearance of the tiny split between the propagation eigenstates ($K_L, K_S$) and the consequent $K_0-\bar K_0$ strangeness oscillations (see also \cite{Feynman_fundamental}). The strong interactions are associated with characteristic lifetimes of $10^{-23}\ \rm{s}$, while weak interactions are associated with characteristic lifetimes of about $10^{-8}\ \rm{s}$. Consequently, the probability that weak interactions affect the production of a particle by strong interactions is indeed negligible.

We shall use a similar argument in the case of neutrinos.  Lifetime is defined as the inverse of the decay rate for an unstable particle. Similarly, we shall define the characteristic time $\tau_m$ as the inverse of the interaction rate for mass-generating interactions \eqref{interaction_rate}:
\begin{equation}\label{tau}
\tau_m=\frac{1}{\Gamma_m}.
\end{equation}
The interaction rate depends on the normalization volume $V$, therefore it is not really an observable, but with a reasonable choice of $V$, it can provide a sensible result and term of comparison.

The normalization volume is customarily taken to infinity, since its ``boundary'' represents the space limits at which the interaction drops to zero and particles are represented by asymptotic states. Intuitively, $V$ is physically and not mathematically infinite and, for interactions of very short range, like the weak interactions, the ``infinite'' $V$ can be actually extremely small. The range of the weak interactions is given by the inverse of the $W$-boson mass, according to the Yukawa potential formula
$${\cal V}(r)=-g^2_W\frac{e^{-M_W r}}{r}.$$
At a distance $R_W=1/M_W$, the potential drops by about $1/e$, and this is what we call the range, i.e. the spatial extension within which the weak interactions are significant in strength. If we consider a distance $R=10^2 R_W$, the potential drops by more than  $e^{-100}$, which can be safely considered a vanishing potential, satisfying the condition of ``infinite'' $V$. We shall therefore estimate $V$ to be the volume of a sphere of radius $R$:
\begin{equation}\label{V}
V=\frac{4\pi}{3}R^3=\frac{4\pi}{3}\,\frac{10^6}{M_W^3}.
\end{equation}

The common intuitive description of particle interactions is via the adiabatic hypothesis: the initial particles start infinitely far apart as asymptotic states, essentially eigenstates of the free Hamiltonian. They come close together in an infinite time, they interact in a certain space domain $\Omega$, and then the final states separate in infinite time and space, becoming eventually asymptotic states, when they do not feel the interaction anymore. In this picture, the interaction region is like a black box, about which we do not have any knowledge. As the mass of particles is an effective manifestation of some interaction in all the theories of the Standard Model kind, the region $\Omega$ in which particles interact is the region in which the final particles also have to get their kinematic mass. In other words, the mass-generating interactions do take place with high probability within the region in which particles interact by the other processes which are under study. Normally, we consider implicitly that the mass generation takes place with unit probability, which is a good approximation if the strength of the mass-generating process is bigger than the strength of the studied interaction.

Let us calculate the characteristic time of the mass-generating interaction of neutrinos. Assuming the eigenvalues of the mass matrix to have typical absolute values of the order $m_\nu=0.1\ \rm{eV}$, and considering the volume in which the weak and mass-generating interactions take place as in \eqref{V}, we obtain the interaction rate:
\begin{equation}
    \Gamma_m^\nu=\frac{4}{3}\ 10^6\frac{m_\nu^4}{M_W^3}\sim0.2 \cdot 10^{-39}\ \rm{GeV}.
\end{equation}
Inverting it, we find the characteristic time for the mass generation of neutrinos in a volume one million times bigger than the region in which the weak interaction takes place:
\begin{equation}
    \tau_m^{\nu} \sim 3 \cdot 10^{15}\ \rm{s}\sim 10^{8}\ \rm{years}.
\end{equation}
In other words, neutrinos are enormously ``stable'' as massless particles, and reluctant to acquire kinematical mass during their production in weak interactions. 

We can repeat the whole argument for the case of the other weakly-interacting particles. Among them, the electrons have the smallest mass, $m_e\approx 5\cdot 10^6\ m_\nu$. Consequently, we find the characteristic time for the mass generation of electrons
\begin{equation}
\tau_m^e\approx\left(5\cdot 10^6\right)^{-4}\: \tau_m^\nu\approx 3\cdot 10^{-12}\ \rm{s},
\end{equation}
which is similar to the lower end of the characteristic time for weak interactions, justifying the fact that the electrons and all the other elementary particles are observed to have kinematical masses.

Thus, we can affirm that the mass-generating interactions have an insignificant effect during the weak interaction processes, where neutrinos are created and annihilated, but become significant during the propagation of the neutrinos. This supports the assumption that the neutrinos participate in weak interactions as massless flavour particles, and acquire an effective refractive mass during propagation over macroscopic distances.

We do not presume to have explained the mechanism by which particles get mass or part of their mass by interaction. This mechanism involves a change between orthogonal vacua, with the second vacuum being a condensate of pairs of particles pertaining to the first vacuum \cite{NJL}.  Such subtleties are hardly ever addressed or even considered in the analysis of processes with massive particles. Our argument above, essentially based on the interpretation of kaon mixing extended to neutrinos, indicates that in the circumstances in which two interactions are many orders of magnitude apart in strength, the vacuum of the stronger interaction is not shifted by the weaker one.

\section{Discussion and outlook}\label{sec:discussion}

We have presented a theory for neutrino oscillations in vacuum, where the flavour neutrinos are described as waves of massless particles and the flavour oscillations are explained by the refractive index of the quantum vacuum, which is nondiagonal with respect to the flavour components of the wave.  The effect of a nondiagonal index of refraction for the flavour waves is to rotate the flavour just as an optically active medium rotates the plane of polarization, which is in agreement with Wolfenstein's proposal that flavour oscillations in matter are analogous to optical birefringence \cite{W-biref} (see also \cite{W}).
We propose that flavour oscillations are generally produced in this way, regardless of whether matter is present or not.

\begin{enumerate}
\item We have argued that flavour neutrinos are created and annihilated in weak interactions as massless particles, since the mass-generating interaction for neutrinos is significantly weaker than the weak interaction. Therefore, we treat all the mass and mixing terms in the Lagrangian \eqref{Lmass} as interactions. At the microscopic level, the nondiagonal (flavour-changing) refractive index for the flavour waves is produced by the coherent forward scatterings of massless flavour neutrinos on the vacuum BEH field. The refractive properties of the vacuum, as well as of matter, can be calculated using quantum field theory and the multiple scattering framework.

 We have considered here only Dirac mass terms in the Lagrangian, which come together with a sterile right-handed neutrino field. In our scenario, the sterility of $\nu_{\ell R}$ is indeed perfect, and the corresponding degrees of freedom do not couple ever (before or after diagonalization) to the $W$ and $Z$ bosons. Consequently, right-helicity neutrinos and left-helicity antineutrinos cannot be created in weak interactions.

\item As it was mentioned in the Sect. \ref{sec:intro}, the same theory of oscillations will work also for the case in which Majorana mass terms for neutrinos are engendered by any of the seesaw mechanisms compatible with SM gauge symmetry. Ultimately, they represent also a propagation in the BEH vacuum, though intermediated by some very heavy particles whose fields, in turn, are treated classically. This would give the oscillations that we observe in experiments, with neutrinos whose energy is much lower than the masses of the seesaw particles.

\item We have obtained a universal speed \eqref{vg} for high-energy neutrinos of a given energy in vacuum. The effect of matter is obtained by adding the relevant sources to the averaged wave equation \eqref{waveequation_2}. As an example, in \secref{sec:matter} we considered the flavour oscillations in constant density matter for neutrinos with energies much below the $W$-boson mass scattering on electrons. In that particular case, the speed of neutrinos is not affected by matter due to the linear dependence of the forward scattering amplitude \eqref{forward_ampl.matter} on energy. It remains for the future to study, in the multiple scattered coherent wave framework, the case of variable density and the analogue of adiabatic conversion and the Mikheyev--Smirnov--Wolfenstein effect.

\item The formalism can be extended to the oscillations of massive particles of any spin (mesons or neutrons, for example). The spin has no bearing in this procedure, because the only process involved besides free propagation is coherent scattering, which does not change any kinematic properties of the particles. For relativistic particles, the equation of motion is Klein--Gordon equation, to which we add a mass term. The masses of $K_0$- or $B_0$-mesons, as well as neutrons, are large or very large, therefore these particles are considered to be emitted massive. Then weak interactions or some BSM interactions, in the case of neutrons, mix the particle and antiparticle and produce the tiny shifts in the phases of the wave function, which account for the flavour oscillations. In all cases of particle-antiparticle oscillations, the mixing angle is $\pi/4$, as expected.

In the case of nonrelativistic particle oscillations, we use the Schr\"odinger equation, which is also written as Helmholtz equation, but with different dispersion relation. For example, the analogue of eq. \eqref{waveequation} is:
\begin{equation}\label{waveequation_nr}
\left[\left(\nabla^2 + p^2\right)\delta_{f'f}  +4\pi N F_{f'f}(0)\right]\Psi_{f}({\bf x}, t)= 0,\ \ \ \ \ p^2=2ME,
\end{equation}
where $M$ is the mass of the meson or neutron, $E$ is its nonrelativistic energy, with which it is emitted, $f,f'$ are flavour indices (for particle, antiparticle, and possibly mirror particles) and $F_{f'f}(0)$ is the transition amplitude of the coherent forward scattering produced by the specific interaction which mixes the flavours (for simplicity, we ignore here the imaginary part of the transition amplitude, which is associated with the decays). Supposing that we have only two flavour mixing, i.e. particle and antiparticle, the two eigenvalues of the refractive index will be given by:
\begin{equation}
n_i^2=1+2\pi \frac{N F_i(0)}{ M\,E} ,\ \ \ i=1,2.
\end{equation}
Evanescence can occur if $F_i(0)$ is negative and $N|F_i(0)|/M>E/2\pi$. Since $N|F_i(0)|/M$ is an extremely small number, the evanescence occurs for such low speed, that it is observationally compatible to the rest state.

\item Measurements of the speed of neutrinos have long been considered as a way to detect the effect of neutrino masses. There are mainly two approaches for analyzing the effect of the neutrino mass. We can consider the spread of flight times of neutrinos with different energies, assuming that the neutrinos respect the relativistic dispersion relation, for example from supernovae \cite{Zatsepin}. Alternatively, in order to obtain the deviation of the speed of neutrinos from the speed of light, we may compare the flight times of neutrinos to the flight times of photons originating from the same source \cite{Stodolsky,Longo}. Moreover, several terrestrial reactor or accelerator neutrino experiments for measuring the speed of neutrinos have been carried out. In our theory, flavour neutrinos in vacuum respect the ultrarelativistic dispersion relation with the refractive mass $\overline{m^2}$, and hence the speed of neutrinos is determined by the refractive mass and the energy \eqref{vg}. Thus, the measurements of the speed of high-energy neutrinos can be used to constrain the value of $\overline{m^2}$.

\item In the standard phenomenological treatment of neutrino oscillations, where  the flavour neutrinos are thought to be produced in weak interactions as a linear superposition of states of different masses, considerable efforts have been invested in understanding the decoherence of neutrinos. Since the wave packets that describe the neutrinos of different masses propagate at different speeds, the separation of the wave packets eventually destroys the coherence of the flavour neutrinos and hence the pattern of neutrino oscillations. In our framework, the flavour neutrinos do not lose their coherence during propagation in vacuum. The conservation of coherence is a significant difference compared to the standard description of oscillating neutrinos, which may provide a way to test our proposal. We stress that the distance-dependent and energy-dependent flavour oscillations \eqref{oscillationprobability2} can still be suppressed by other factors, e.g. by averaging over energy or momentum or dissipative interactions producing quantum decoherence (see, for example, \cite{GB}). Indeed, such questions regarding the oscillations of flavour neutrino wave trains and the source of their suppression in experimental settings need to be analyzed further.

\item Since the standard description of neutrinos involves interference of waves of particles of different masses, which is a deviation from standard quantum field theory, several discussions and debates on the description and interpretation of such interference has occurred. As an example, it has been posited that the massive states that describe a flavour neutrino state should have the same energy in order for their interference to be observable \cite{Lipkin} (see also \cite{Stodolsky:wp}), while other authors have held that interference is observable regardless of whether the energies (or momenta) are equal, as the massive neutrinos must be described as wave packets that contain a range of energies \cite{Giunti:wp}. We shall not revisit those discussions here. Instead we note that our proposal shows that massive neutrinos of different masses are not necessary for the description of oscillating neutrinos. We can describe neutrinos as flavour waves which have a common dispersion relation (see \secref{sec:speed}) and which have a distance-dependent and energy-dependent envelope.

\item Experiments for a kinematic determination of the neutrino mass are conducted using a $\beta$-decay or an electron capture process of nuclei with preferably a low energy release ($Q$-value). Since in such processes the produced neutrino takes away at least the energy stored in its kinematic mass, the high-end of the electron energy spectrum would be affected by the neutrino mass. Currently, the most sensitive experiment of this kind is KATRIN, where an upper bound on the kinematic mass of electron neutrino has been obtained as $m_{\nu_e} < 0.45\;\mathrm{eV}$ \cite{KATRIN:2021uub} (with the projected final sensitivity of the experiment in the vicinity of $0.2\;\mathrm{eV}$). Since in our proposal flavour neutrinos are produced in weak interactions as massless particles and subsequently gain their refractive mass due to the interactions with the vacuum BEH field (and matter if it is present), the effect of the universal refractive mass of neutrinos can only be observed in experiments where neutrinos are detected after they have propagated for some considerable macroscopic distance. When the consequences of our framework for low energy neutrinos are understood, experiments like KATRIN, which probe the creation of low energy neutrinos, can be used to test our proposal. Our prediction is that the kinematical flavour neutrino mass in weak interaction will be zero, and this is compatible with any upper bound that KATRIN might produce.

\item By extrapolating the results for the refractive index to the very low energy, we have obtained an evanescent behaviour of neutrinos. This is a direct effect of the fact that we worked with the proper relativistic wave equation, which always leads to a formula for $n^2$, and not for $n$. The same is true also for massive nonrelativistic waves treated by Schr\"odinger's equation, which becomes the Helmholtz equation in the momentum space \cite{Lax, Foldy}. The evanescence of quantum mechanical probability waves is related to the tunnel effect.  Since the energies of the neutrinos where evanescence appears are so small, it is unlikely that any terrestrial experiment could test this effect, possibly with the exception of PTOLEMY. However, very low energy neutrinos have huge implications in structure formation and other cosmological phenomena \cite{Dolgov}. Also in very dense matter, the evanescence cutoff energy can change a lot. These aspects need to be further studied.

\end{enumerate}

\section*{Acknowledgments}

We are most grateful to Iver Brevik,   Kazuo Fujikawa, Dieter Haidt, Mikhail Vysotsky, Jenny Wagner, and in particular to Masud Chaichian, for  illuminating discussions and insightful comments. NS acknowledges the financial support from the Vilho, Yrj\"o and Kalle V\"ais\"al\"a Foundation.

\appendix

\section{Forward scattering amplitude in BEH vacuum}\label{sec:forwardscattering}

The scattering amplitude $f_{\ell' \ell}(0)$ for the forward scattering process $\nu_{\ell L}\to\nu_{\ell'L}$ is related to the differential cross section of interaction by the formula
\begin{equation}\label{scatteringamplitude}
    |f_{\ell' \ell}(0)| = \sqrt{\frac{d \sigma_{\ell \rightarrow\ell'}}{d \Omega}\Big|_{\theta,\phi=0}}.
\end{equation}
Hence $f_{\ell' \ell}(0)$ can be calculated from quantum field theory up to a phase. Treating \eqref{Lmass} as an interaction leads to invariant transition matrix elements between flavour neutrinos (for details of calculation, see \cite{AT2023}). To lowest order, these are
\begin{align}\label{invarianmatrixelements}
    i \mathcal{M}_{\nu_{\ell L} \rightarrow \nu_{\ell' L}} & = \sum_{\ell''} \; \feynmandiagram [inline=(a.base), layered layout, horizontal=a to b] {
  a -- [fermion, edge label=\(\nu_{\ell L}\), momentum'=\(p\)] b -- [fermion, insertion=0.01, insertion=0.99, edge label=\(\nu_{\ell'' R}\), momentum'=\(p\)] c -- [fermion, edge label=\(\nu_{\ell' L}\), momentum'=\(p\)] d, 
}; 
= - 2i \left( M M^\dagger \right)_{\ell' \ell},  \notag \\
i \mathcal{M}_{\Bar{\nu}_{\ell R} \rightarrow \Bar{\nu}_{\ell' R}} & = \sum_{\ell''} \; \feynmandiagram [inline=(a.base), layered layout, horizontal=a to b] {
  a -- [anti fermion, edge label=\(\Bar{\nu}_{\ell R}\), momentum'=\(p\)] b -- [anti fermion, insertion=0.01, insertion=0.99, edge label=\(\Bar{\nu}_{\ell''L}\), momentum'=\(p\)] c -- [anti fermion, edge label=\(\Bar{\nu}_{\ell' R}\), momentum'=\(p\)] d, 
}; 
= - 2i \left( M M^\dagger \right)_{\ell' \ell}^* ,
\end{align}
where $L$ and $R$ denote helicities.
The 1-particle-to-1-particle differential scattering cross section of a massless particle in a normalization volume $V$ and during a very long but finite time interval $T$ is
\begin{align}\label{differentialcrosssection}
    \frac{d\sigma}{d \Omega} &=\frac{V}{T}\frac{d}{d \Omega}\int\frac{V\,d^3\mathbf{p}'}{(2\pi)^3} (2\pi)^4 \delta^4(p' -p) VT |\mathcal{M}|^2\frac{1}{2EV}\frac{1}{2E'V} \notag \\
    & = V \int  dE' E'^2 (2\pi) \delta^4(p' -p) |\mathcal{M}|^2\frac{1}{4E'E} \notag \\
    & = \left. \frac{\pi }{2} V\delta^3(\mathbf{p}'-\mathbf{p}) |\mathcal{M}|^2 \right|_{|\mathbf{p}'| = |\mathbf{p}|},
\end{align}
where $E=|\mathbf{p}|$, $E'=|\mathbf{p}'|$ and the incident neutrino flux is $1/V$ (massless neutrinos propagate with the speed of light).
Using formulas \eqref{scatteringamplitude}, \eqref{invarianmatrixelements}, \eqref{differentialcrosssection} and $\delta^3(0) = \frac{V}{(2\pi)^3}$ leads to the following result for the forward transition amplitudes:
\begin{align}\label{forward_ampl}
    f_{\ell' \ell}(0) &= - \frac{V}{2\pi} \left(M M^\dagger\right)_{\ell' \ell}, \ \ \ \ \ \ \ \mbox{for neutrinos},\notag\\
    \bar f_{\ell' \ell}(0) &= - \frac{V}{2\pi} \left(M M^\dagger\right)^*_{\ell' \ell}, \ \ \ \ \ \ \ \mbox{for antineutrinos},
\end{align}
with the phase determined in Appendix \ref{app_phase}.

Incidentally, we can also define an interaction rate, namely the probability of interaction per unit time:
\begin{align}\label{interaction_rate}
\Gamma_m^{{\nu_\ell}\rightarrow{\nu_{\ell'}}} &=\frac{1}{T}\int\frac{V\,d^3\mathbf{p}'}{(2\pi)^3} (2\pi)^4 \delta^4(p' -p) VT |\mathcal{M}|^2\frac{1}{2EV}\frac{1}{2E'V} \notag \\
    & = \left. 2\pi^2 \delta^3(\mathbf{p}'-\mathbf{p}) |\mathcal{M}|^2 \right|_{|\mathbf{p}'| = |\mathbf{p}|}\notag \\
    &=\frac{V}{\pi}\left|(MM^\dagger)_{\ell' \ell}\right|^2.
\end{align}
This has the dimension of a decay rate and is used in Sect. \ref{sec:masslessness} to define the characteristic time of mass-generating interactions.

\section{Phase of the forward scattering amplitude}\label{app_phase}

The phase of the forward scattering amplitude $f_{\ell' \ell}(0)$ is determined by comparing the $S$-matrices of quantum mechanics and quantum field theory.
The quantum mechanical $S$-matrix is
\begin{equation}\label{S-matrix.QM}
    \bra{f}S\ket{i}=\delta_{fi}-2\pi i\delta(E_f-E_i)T_{fi},\qquad T_{fi}=\bra{f}T\ket{i},
\end{equation}
where $\ket{i}$ and $\ket{f}$ are the initial and final free particle states, $E_i$ and $E_f$ are the corresponding energy eigenvalues, and $T_{fi}$ is the transition matrix. In the first Born approximation, corresponding to first order perturbation theory, the transition operator $T$ is equal to the interaction Hamiltonian (i.e. the classical potential).

In quantum field theory, we define the $S$-matrix in terms of the Feynman amplitude $\mathcal{M}_{fi}$ as
\begin{equation}
    \bra{f}S\ket{i}=\delta_{fi}+(2\pi)^4\delta^4(p_f-p_i)\frac{1}{\sqrt{2VE_{\mathbf{p}_f}}}\frac{1}{\sqrt{2VE_{\mathbf{p}_i}}}i\mathcal{M}_{fi}.
\end{equation}
For coherent forward scattering with $p_f=p_i=(E_\mathbf{p},\mathbf{p})$, $\delta^3(\mathbf{p}-\mathbf{p}) = \frac{V}{(2\pi)^3}$, we have
\begin{equation}\label{S-matrix.QFT}
    \bra{f}S\ket{i}=\delta_{fi}+2\pi i\delta(E_\mathbf{p}-E_\mathbf{p})\frac{\mathcal{M}_{fi}}{2E_\mathbf{p}}.
\end{equation}
Therefore, from the comparison of \eqref{S-matrix.QM} and \eqref{S-matrix.QFT}, we have the correspondence
\begin{equation}\label{amplitudecorrespondence}
    T_{fi}=-\frac{\mathcal{M}_{fi}}{2E_\mathbf{p}}.
\end{equation}

In the multiple scattering theory \cite{Lax}, the averaged transition operator $\bar T$ enters the Helmholtz equation for averaged wave $\Psi(\mathbf{x})$ as 
\begin{equation}
    \left(\nabla^2 +\mathbf{p}^2 -C\bar T\right)\Psi(\mathbf{x})=0,
\end{equation}
where $C$ is a positive parameter.\footnote{In Ref. \cite{Lax}, this is derived in the context of the Schr\"odinger equation for nonrelativistic particles, but the discussion is actually the same for massless particles, like the photon, as shown in Ref. \cite{Born} (Chapter 13), because the Helmholtz equation is formally the same and the conditions of applicability are in both cases similar, namely those for the validity of the first Born approximation.}
Comparison with the wave equation \eqref{waveequation} gives that the forward scattering amplitude $f(0)$ is equal to $-\bar T$ up to a positive factor (of mass dimension $-2$),
\begin{equation}\label{f0)-T-relation}
    4\pi N f(0)=-C\bar T.
\end{equation}
In our case, the transition matrix for neutrinos is given by \eqref{invarianmatrixelements} and \eqref{amplitudecorrespondence} as
\begin{equation}\label{transitionmatrix}
    T_{\ell'\ell}=-\frac{1}{2E_\mathbf{p}}\mathcal{M}_{\nu_{\ell L} \rightarrow \nu_{\ell' L}}
    =\frac{1}{E_\mathbf{p}}\left( M M^\dagger \right)_{\ell' \ell},
\end{equation}
and its relation to the forward scattering amplitude is given by \eqref{f0)-T-relation} as $f_{\ell' \ell}(0) = -\frac{C}{4\pi N} T_{\ell'\ell}$. Since the absolute value of the forward scattering amplitude is given by \eqref{scatteringamplitude}, and with the number density of scatterers taken as $N=\frac{1}{V}$, we have $C=2E_\mathbf{p}$.
Thus, we obtain that the forward scattering amplitude has opposite sign compared to the transition matrix \eqref{transitionmatrix}:
\begin{equation}
    f_{\ell' \ell}(0)=-\frac{VE_\mathbf{p}}{2\pi}\,T_{\ell'\ell}
    =-\frac{V}{2\pi}\left( M M^\dagger \right)_{\ell' \ell}.
\end{equation}

\end{document}